\documentclass[prl, twocolumn, footinbib,a4paper, showpacs]{revtex4}
\usepackage[latin1]{inputenc}
\usepackage{amssymb}
\usepackage{amsmath}
\usepackage{amsbsy}
\usepackage{bm}
\usepackage{graphicx}

\begin{document}
\title{On the role of step-flow advection during electromigration-induced step bunching}
\author{Matthieu Dufay $^1$,  Thomas  Frisch $^2$, and  Jean-Marc Debierre$^1$}
\affiliation{$^{1}$ Laboratoire Mat\'eriaux et Micro\'electronique
de Provence, Aix-Marseille Universit\'e and CNRS, Facult\'e des
Sciences et Techniques de Saint-J\'er\^ome, Case 151, 13397
Marseille Cedex 20, FRANCE \\
$^{2}$ Institut de Recherche sur les
Ph\'enom\`enes Hors Equilibre,
Aix-Marseille Universit\'e, Ecole Centrale Marseille, and CNRS,\\
49, rue Joliot Curie, BP 146, 13384 Marseille Cedex 13, FRANCE}

\begin{abstract}
We propose a one-dimensional model based on the Burton-Cabrera-Frank
equations to describe the electromigration-induced step bunching
instability on vicinal surfaces. The step drift resulting from
atomic evaporation and/or deposition is explicitly included in our
model. A linear stability analysis reveals several stability
inversions as the evaporation rate varies, while a deposition flux
is shown to have a stabilizing effect.
\end{abstract}


\maketitle

Due to its importance for both fundamental science and technological
applications in microelectronics, research in the field of crystal
growth on semiconductor vicinal surfaces has become increasingly
active \cite{saito98,pimpinelli98,jeong99, politi00,
yagi01,stangl04}. Vicinal surfaces are slightly misoriented with
respect to closed-packed crystalline planes of low indices, such as
the well-studied Si(111) plane. The atomic steps resulting from the
miscut are generally mobile (step flow) and subject to a number of
instabilities as step bunching or meandering \cite{liu98bis,
jeong99, frisch05, sato05, pierre-louis06, dufay07}. At long times,
nonlinear dynamics lead to the formation of more or less
well-organized surface patterns. The idea to control these patterns
is appealing, a major issue being the self-organization of
nano-structures on semiconductor surfaces. A few ways to reach this
goal are currently under investigation, such as imposing a net
atomic flux (evaporation/deposition), using elastic stresses or
applying a constant electrical field inducing an adatom drift
(surface electromigration). Combinations of these methods allow
extra flexibility \cite{stangl04, fruchart05}. Surface
electromigration was first observed by Latyshev and co-authors
\cite{latyshev89}, and an early theory proposed by Stoyanov
\cite{stoyanov91}. The stability of a Si(111) vicinal surface may
change according to the temperature, the sign of the net  atomic
flux, and the direction of the electrical current. Three temperature
regimes have been identified experimentally \cite{metois99,
fujita99,gibbons05,gibbons06}. Both in the high and low  temperature
regimes, the step bunching instability appears when the electrical
current  is applied in the step-down direction. Alternatively, in
the intermediate regime, it may be necessary (according to the
experimental conditions) to apply a step-up current to trigger step
bunching. This  stability inversion with respect to the direction of
the electrical current is still the subject of active
research and  different mechanisms  have been proposed such as
sign variations of  the effective charge number ($Z^*$) with temperature,
step transparency \cite{pierre-louis03, pierre-louis04}, and two-region terrace diffusion
\cite{zhao04}. While  sign variations of $Z^*$ with temperature have been ruled
out  by a recent experiment of mass transport across a trench
\cite{degawa00}, the two other possibilities remain. However, due to
the large number of experimental parameters and ultra-high vacuum
conditions, direct in-situ experimental measurements of adatom
diffusion at the step edges are still  difficult to achieve. In this
Letter, we propose a one-dimensional model based on the
Burton-Cabrera-Frank equations \cite{Burton51} to describe the step
flow instabilities arising when electromigration is combined with a
net atomic flux. We first remark that, under typical experimental
conditions, the advection effects due to the mean step velocity are
comparable in magnitude with the drift due to the electromigration
current. Both effects are thus included in our model and a linear
stability analysis shows that the interplay between them does
provoke stability inversions as evaporation is increased.

\begin{figure}[t]
\includegraphics[width=0.4\textwidth]{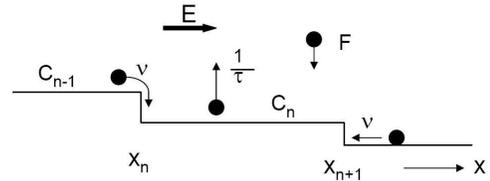}
\vskip-20pt
\caption{Schematic side view of a vicinal surface: notations and definitions.
The electrical field is imposed in the step-down direction here.
\label{Fig:schema} }
\end{figure}
During evaporation or growth, the steps of Si(111) vicinal surfaces undergo
a global drift with a  constant mean velocity $V_0$. In the frame moving at this velocity,
the adatom concentration $C_n$ on the $n$-th terrace obeys the following quasi-static
equation,
\begin{equation}
D_s \partial_{XX} C_n +\Big(V_0- \frac{D_s}{\ell_E}\Big)
\partial_{X} C_n - \frac{C_n}{\tau} + F = 0,
\label{Eq:Diffusion}
\end{equation}
where $D_s$ is the diffusion coefficient, $\ell_E=k_B T/(Z^*eE)$
the electrical length, $\tau$ the desorption time, and $F$ the atomic
deposition flux. As shown in Fig. (\ref{Fig:schema}), the electrical field
reads $\mathbf{E}=E \mathbf{\hat X}$, with $\mathbf{\hat X}$ the unit vector
pointing in the step-down direction.
At terrace ends $X=X_n$ and  $X=X_{n+1}$, Eq. (\ref{Eq:Diffusion}) is subject to
the following boundary conditions,
\begin{eqnarray}
( D_s \partial_X + V_0 - D_s/\ell_E) C_n &=& \nu [C_n-C_n^{eq}], \\
( D_s \partial_X + V_0 - D_s/\ell_E) C_n &=& -\nu [C_n-C_{n+1}^{eq}],
\label{Eq:Boundary}
\end{eqnarray}
which ensure mass conservation. Here $\nu$ is the step kinetic
coefficient, $C_n^{eq}$ the adatom equilibrium concentration at step
$n$, and we neglect step transparency. We can remark from the
structure of Eqs. (\ref{Eq:Diffusion}-\ref{Eq:Boundary}) that the
advection and the electromigration terms play a similar role. The
associated velocities $V_0$ and $D_s/\ell_E$  may be both positive
or negative, according to the sign of the imposed net atomic flux and
electrical current. The advection terms are usually neglected in
theoretical studies, which is equivalent to setting $V_0$ to zero in
Eqs. (\ref{Eq:Diffusion}-\ref{Eq:Boundary}). However, on the basis
of the present experimental knowledge, it seems unrealistic to
neglect advection as compared to electromigration. On one hand, the
net atomic flux varies in a rather wide range in practice (three
decades) while the mean terrace width extends over one decade at least.
As a consequence, the corresponding values of the mean step velocity
$V_0$ typically range from $10^{-10} $ to $10^{-6}$ m.s$^{-1}$. On
the other hand,  due to the experimental uncertainties on the values
of the effective charge number $Z^*$ and, to a less extent, to the
temperature influence on the adatom diffusion coefficient $D_s$,
estimates of the drift velocity $D_s/\ell_E$ are found to lie in the
very same range. Excepted in equilibrium conditions, the two effects
are thus similar both in nature and in magnitude, so that we will
keep the $V_0$ terms in our model.
At a given time $t$, the velocity
of step $n$ is $V_n= \Omega_s \nu [C_n+C_{n-1}-2 C^{eq}_n] -V_0$,
where the concentrations are evaluated at $X=X_n(t)$, and $\Omega_s$
represents the unit atomic surface. The terrace lengths
$L_n(t)=X_{n+1}(t)-X_n(t)$ vary slowly in time, with a constant mean
value $L_0$. We introduce standard step-step repulsive interactions
in the adatom equilibrium concentration, $C_n^{eq}/C_0= 1 + A
(1/L_n^3-1/L_{n-1}^3)$,
where $k_BT A\Omega_s^{-1}$ is the step-step interaction coefficient \cite{muller04}.
Setting the unit length to the initial terrace width $L_0$ and the unit time to
$t_0=L_0^2/D_s$, we define the dimensionless variables $x=X/L_0$, $v_0=V_0
t_0/L_0$, $c_n=C_n/C_0$, and $c_n^{eq}=C_n^{eq}/C_0$. For these
variables, the previous equations become
\begin{eqnarray}
(\partial_{x} + v_0 - \eta) \partial_x c_n - s^2 c_n +f =0,& x_n<x< x_{n+1},
\label{Eq:BCF}\\
 (\partial_x+v_0-\eta)\ c_n =\rho (c_n-c_n^{eq}) , &\quad x=x_n,\\
(\partial_x+v_0-\eta)\ c_n =-\rho (c_n-c_{n+1}^{eq}), &\quad  x=x_{n+1},\\
v_n = \phi \rho (c_n+c_{n-1}-2 c_n^{eq})-v_0, &\quad x=x_n,\end{eqnarray}
with $c_n^{eq}=1+a(1/l_n^3-1/l_{n-1}^3)$. The system dynamics is
thus controlled by {\em six} independent nondimensional parameters:
$\eta=L_0/\ell_E$ is proportional to the electrical field,
$s^2=L_0^2/(D_s \tau)$ involves the rate of desorption and $f=F
L_0^2 /(D_s C_0)$ the atomic deposition flux, $\rho=\nu L_0/D_s$ compares
attachment to diffusion, $\phi=\Omega_s C_0$ gives the proportion of
occupied sites, and $a=A/L_0^3$ measures step-step repulsion.
These six parameters are not completely independent in practice.
As a consequence, when temperature
is varied in a given experimental setup, the system is driven along a rather
complex trajectory in the parameter space. This trajectory is very likely
to cross a succession of stable and unstable regions, as often observed in practice.

\begin{figure}[t]
\includegraphics[width=0.3\textwidth]{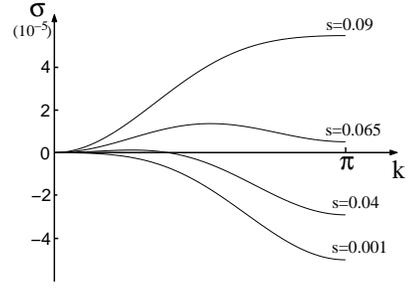}
\caption{Growth rate $\sigma$
as a function of the wave number $k$, as obtained from Eqs. (\ref{Eq:sigmas}, \ref{Eq:gr}).
Different evaporation rates $s$ are considered.
 Nondimensional parameters are $\phi=0.2$, $f=0$,
$\eta=-10^{-4}$, $\rho=20.0$, $a=5\times 10^{-6}$.
 \label{Fig:grate} }
\end{figure}
To provide quantitative support to this idea, we now perform the
linear stability analysis of a uniform train of steps traveling at a
constant velocity $v_0$ in the laboratory frame. Let us remark that
we keep the notation $t$ for the dimensionless time hereafter. A
general solution of Eq. (\ref{Eq:BCF}) giving the concentration
profile on terrace $n$ is $c_n=f/s^2+\alpha_n e^{r_1 x}+\beta_n
e^{r_2 x}$, where $r_{1,2} = -(v_0-\eta)/2 \pm
1/2[(v_0-\eta)^2+4s^2]^{1/2}$. To avoid lengthy expressions, we
first set $f=0$ and we discuss the case $f>0$ later on. For the
unperturbed system, we find that $v_0$ obeys the following
transcendental equation
\begin{equation}
\frac{v_0}{\phi \rho}= \frac{2s^2(u_1-u_2)-\rho
u_1u_2\Lambda}{d_1-d_2},
\label{Eq:v0}
\end{equation}
where $\Lambda=r_1-r_2$, $u_1=e^{r_1}-1$, $u_2=e^{r_2}-1$,
$d_1=(r_1+\rho)(r_2-\rho)e^{r_1}$, and
$d_2=(r_1-\rho)(r_2+\rho)e^{r_2}$.
The linear stability of the system is tested by adding a small harmonic perturbation
of wave number $k$, on the step positions, $x_n=n+\epsilon e^{ikn+\xi t}$. In order to find
the dispersion relation $\xi=\sigma+i \omega$, we expand the step
velocities up to the first order in the perturbation amplitude, $\epsilon\ll1$. We obtain
the following exact expression for the growth rate:
\begin{equation}
\sigma (k)= 2 \phi \rho \sin^2{\Big(\frac{k}{2}\Big)} \
\frac{A_1 s^2+ a (A_2 + A_3 \cos{k})}{s^2(d_1-d_2)^2},
\label{Eq:sigma}
\end{equation}
where
\begin{eqnarray}
\frac{A_1}{\Lambda \rho}&=&\big[r_1d_2-r_2d_1+2\rho \Lambda
(r_1+r_2)\big]e^{r_1+r_2}+r_2d_2-r_1d_1,
\nonumber \\
A_2&=&6 s^2(d_2-d_1)\big[2s^2(u_2-u_1)-\rho \Lambda (u_1+u_2+2)\big],
\nonumber \\
A_3&=& 6 s^2 \rho \Lambda(d_2-d_1)(1+e^{r_1+r_2}).
\end{eqnarray}
In the parameter space, this allows us to compute the critical values of the parameters
defining the boundaries between stable and unstable regions.
One thus has to solve the system of two equations, Eqs. (\ref{Eq:v0}-\ref{Eq:sigma}),
in the two unknowns, $\eta$ and $v_0$.
Remarking that the condition $\vert v_0-\eta \vert \ll s$ is always verified
for the practical values of the electrical field and net atomic flux,
we obtain analytical expressions for the velocity,
\begin{equation}
v_0=-2 \phi \rho s \ \frac{e^s-1}{\rho(1+e^s)+s(e^s-1)},
\label{Eq:v0s}
\end{equation}
and the growth rate,
\begin{equation}
\frac{\sigma (k)}{4 \phi \rho}=\sin^2{\Big(\frac{k}{2}\Big)}
\frac{B_1 (v_0-\eta)+a (B_2+ B_3 \cos{k})}{s \ d_3^2},
\label{Eq:sigmas}
\end{equation}
with $d_3=(s+\rho)^2 e^{2s}-(s-\rho)^2$, and
\begin{eqnarray}
&&\frac{B_1}{\rho s^2 e^s}= (s+\rho)^2 (s-1) \ e^{2s} - 4 \rho s\ e^{s} + (s-\rho)^2 (s+1),
\nonumber \\
&&B_2=-6 s^2 d_3 \big[(s+\rho) e^{2s} +\rho-s \big], B_3= 12 s^2 d_3 \rho \ e^s.
\label{Eq:gr}
\end{eqnarray}
\begin{figure}[t]
\includegraphics[width=0.3\textwidth]{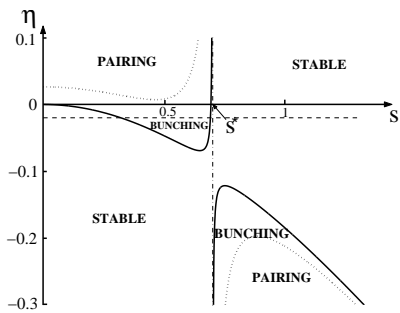}
\caption{Stability diagram in the $(s, \eta)$ plane.
Solid (dotted) curves represent $\eta_c$ ($\eta_p$).
Nondimensional parameters are $\phi=0.2$, $f=0$,
$\rho=20.0$, $a=10^{-4}$. For a step-up electrical field,
the horizontal dashed line represents a trajectory as the
evaporation rate increases.
\label{Fig:diagstab} }
\end{figure}
Depending on the physical parameters, the growth rate $\sigma (k)$
assumes negative or positive values. In the latter case, the one-dimensional
train of steps is linearly unstable.
The variations of the growth rate with the perturbation wave number $k$ are displayed
in Fig. (\ref{Fig:grate}) for different values of the evaporation parameter $s$.
The electrical field is imposed in the step-up direction here ($\eta<0$). For $s=0.001$
all the modes are stable, while the large wavelengths are unstable
for $s=0.04$. At higher evaporation rates, all the wave numbers are
unstable. Note that for $s=0.09$, the most unstable mode is obtained for $k=\pi$
(step-pairing instability).
Using Eqs. (\ref{Eq:v0s},\ref{Eq:sigmas},\ref{Eq:gr}), we obtain
the analytical expression for the boundary lines $\eta_c(s)$
separating the stable and the step bunching
regions in the $(s, \eta)$ plane:
\begin{equation}
\eta_c=v_0+\Big(\frac{B_2+B_3}{B_1}\Big) a.
\label{Eq:etac}
\end{equation}
\begin{figure}[t]
\includegraphics[width=0.3\textwidth]{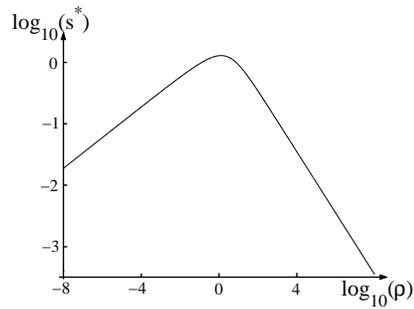}
\caption{ Critical value of the nondimensional
evaporation rate $s^*$ as a function of the
attachment/diffusion parameter $\rho$ for $f=0$
(log-log scales).
\label{Fig:sstar} }
\end{figure}
In addition, one shows that step-pairing instability becomes the most unstable
mode for $\eta_p=v_0+a(B_2-3 B_3)/B_1$. The stability diagram
representing the boundaries between the stable, and unstable regions
is shown in Fig. (\ref{Fig:diagstab}). For typical values of the
step-up electrical field ($500 \rm{V/m}$), $\eta$ assumes small
negative values (lower than $10^{-2}$ in magnitude) which lie
slightly below the horizontal axis in the $(s, \eta)$ plane
\footnote{Under typical  experimental conditions on a
Si(111) surface, $E=400-700$Vm$^{-1}$, $L_0=10^{-9}-10^{-5}$m,
$A=10^{-28}-10^{-26}$m$^3$,
$T=1200-1650$K, $Z^*=10^{-3}-10^{-1}$,
$C_0 \Omega_s=0.2$, $\tau=\tau_0 \exp(E_{des}/k_b T)$, with
$\tau_0=10^{-15}$s, $E_{des}=4.2$eV, $D_s=D_0\exp(-E_d/k_b T)$,
with $D_0= 10^{-6}$ m$^2/$s, $E_d=1.1$eV. We find for instance $s=10^{-4}-10^0$
and $\eta=10^{-2}-10^{-8}$.
}.
Since the evaporation rate $s$ becomes larger as temperature is increased,
the system is stable both at low and high temperatures and it
becomes unstable in the intermediate temperature range. Indeed, this
succession of stability reversals is observed experimentally for
step-up currents \cite{metois99, fujita99, gibbons05, gibbons06}. In
the unstable regions, both for step-up and step-down currents, step
bunching is superseded by the step-pairing instability for large
electrical fields. Moreover,  the bunching regions progressively
reduce in size as the step-step interaction coefficient $a$ is
decreased.
As shown by Eqs. (\ref{Eq:gr}, \ref{Eq:etac}), the location of  the
vertical asymptote $s=s^*$ solely varies with the
attachment/diffusion parameter $\rho$. These variations are
represented in Fig. (\ref{Fig:sstar}).
\begin{figure}[t]
\includegraphics[width=0.4\textwidth]{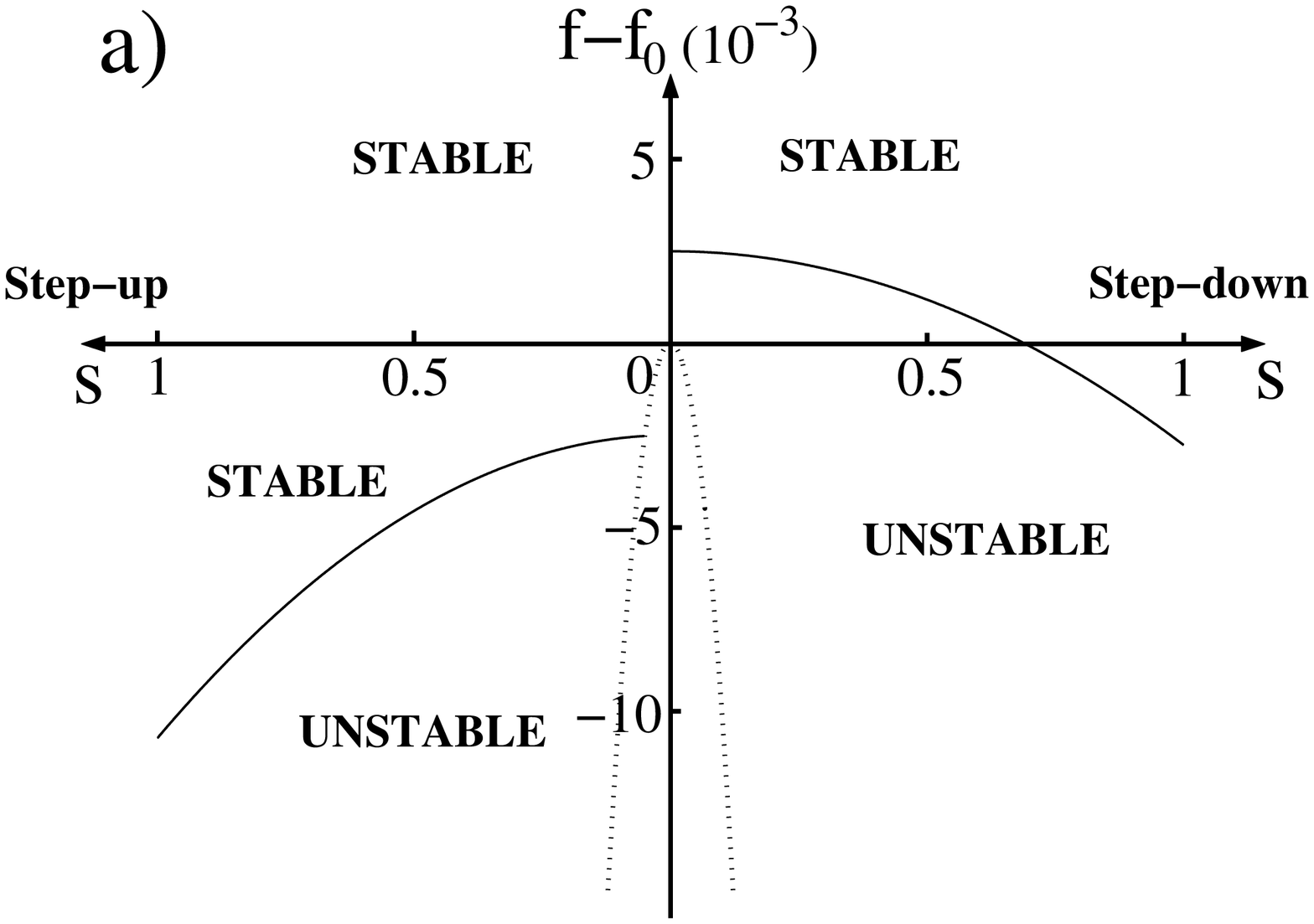}
\vskip20pt
\includegraphics[width=0.4\textwidth]{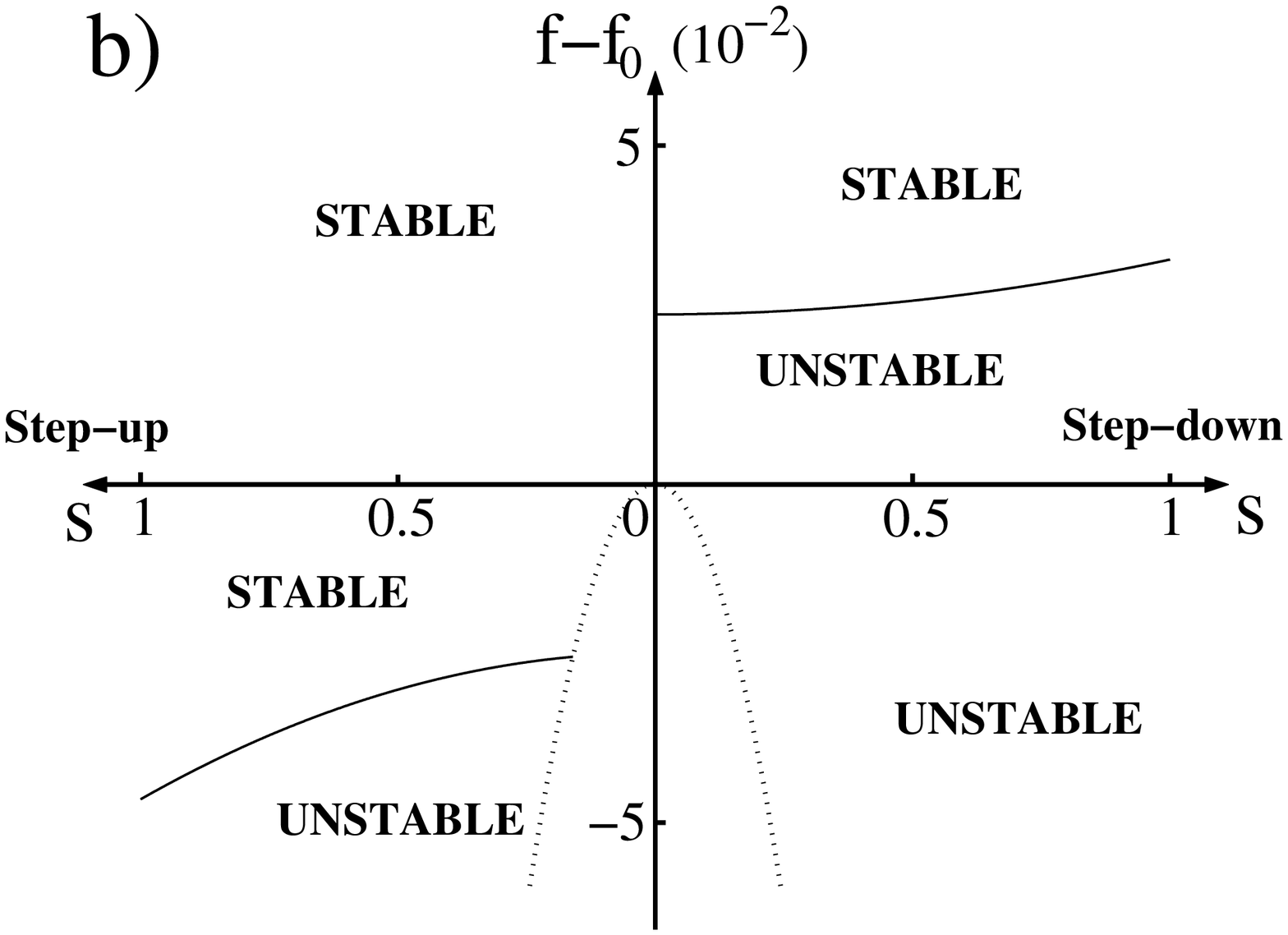}
\caption{ Stability diagrams in the $(s, f-f_0)$ plane.
Horizontal axes $f=f_0$ correspond to zero net flux.
On the left (right) side of the vertical axes, the electrical
current is step-up (step-down).
Solid lines represent the boundaries between stable and
unstable regions. The regions below the dotted lines
are unphysical ($f<0$).
Nondimensional parameters are $\phi=0.2$,
$\rho=1.0$, $a=10^{-4}$, and we set
a) $\eta=\pm 5.0\times 10^{-4}$, b) $\eta=\pm 5.0 \times 10^{-3}$.
\label{Fig:flux}}
\end{figure}
A maximum is obtained for
$\rho \simeq 1$, when the attachment length $d=D_s/\nu$ is
comparable to the terrace width $L_0$. For attachment-limited
dynamics,  $\rho<1$, we find that $s^*\simeq (12\rho)^{1/4}$, while
$s^*\simeq (12/\rho)^{1/2}$ for diffusion-limited dynamics, $\rho>1$.
As discussed above, $\vert \eta \vert<10^{-2}$ in practice, so that
the higher critical value of the evaporation rate is very close to
$s=s^*$. Since the desorption time reads $\tau=L_0^2/(s^2 D_s)$, we
respectively obtain $\tau^*=(L_0^3/12\nu D_s)^{1/2}$, and
$\tau^*=\nu L_0^3/12 D_s^2$ in the attachment-limited and the
diffusion-limited regimes. This implies that $s^*$ strongly depends
on the miscut angle.
Let us discuss now the case of a nonzero
deposition flux, $f> 0$. The drift velocity reads
\begin{equation}
v_0=-2 \phi \rho \ \frac{(s^2-f)(e^s-1)}{s \rho(1+e^s)+s^2 (e^s-1)}.
\label{Eq:v0f}
\end{equation}
It vanishes when the deposition and evaporation fluxes compensate,
$f=f_0=s^2$. We first consider the case of a step-up current.
The step bunching instability occurs only when evaporation dominates deposition,
$f<f_0$, as shown on the left side of Fig. (\ref{Fig:flux}).
On the other hand, for a step-down current, the instability may occur
both for $f<f_0$ and $f>f_0$, depending on the values of the physical
parameters. From  Fig. (\ref{Fig:flux}), it is also clear that increasing the deposition
flux has a stabilizing effect, whatever the direction of the electrical field.
Since $a\sim L_0^{-3}$, after Eq. (\ref{Eq:etac}) the stability threshold
$\eta_c\simeq v_0$ for large $L_0$ values. Thus, measurements of the step velocity
$V_0$ could provide direct estimates of the effective charge number,
$Z^*=(V_0k_B T)/(D_s e E)$.

In conclusion, we have shown that advection can not in general be
neglected as compared to electromigration.  We have identified a set of six
nondimensional parameters governing the system evolution. For
step-up electrical currents, two stability inversions have been
found as evaporation increases. Numerical simulations of our model
equations are currently performed to investigate nonlinear regimes,
such as coarsening dynamics, oscillating modes and spatiotemporal
chaos \cite{krug05, popkov06, chang06}.

\acknowledgments{It is a pleasure to thank F. Leroy, J.-J. M\' etois, P. M\H uller, O. Pierre-Louis, and A. Verga for fruitful discussions.}

\bibliography{advect.bib}

\begin{thebibliography}{27}
\expandafter\ifx\csname natexlab\endcsname\relax\def\natexlab#1{#1}\fi
\expandafter\ifx\csname bibnamefont\endcsname\relax
  \def\bibnamefont#1{#1}\fi
\expandafter\ifx\csname bibfnamefont\endcsname\relax
  \def\bibfnamefont#1{#1}\fi
\expandafter\ifx\csname citenamefont\endcsname\relax
  \def\citenamefont#1{#1}\fi
\expandafter\ifx\csname url\endcsname\relax
  \def\url#1{\texttt{#1}}\fi
\expandafter\ifx\csname urlprefix\endcsname\relax\def\urlprefix{URL }\fi
\providecommand{\bibinfo}[2]{#2}
\providecommand{\eprint}[2][]{\url{#2}}

\bibitem[{\citenamefont{Saito}(1998)}]{saito98}
\bibinfo{author}{\bibfnamefont{Y.}~\bibnamefont{Saito}},
  \emph{\bibinfo{title}{Statistical Physics of Crystal Growth}}
  (\bibinfo{publisher}{World Scientific}, \bibinfo{year}{1998}).

\bibitem[{\citenamefont{Pimpinelli and Villain}(1998)}]{pimpinelli98}
\bibinfo{author}{\bibfnamefont{A.}~\bibnamefont{Pimpinelli}} \bibnamefont{and}
  \bibinfo{author}{\bibfnamefont{J.}~\bibnamefont{Villain}},
  \emph{\bibinfo{title}{Physics of Crystal Growth}}
  (\bibinfo{publisher}{Cambridge University Press}, \bibinfo{year}{1998}).

\bibitem[{\citenamefont{Jeong and Williams}(1999)}]{jeong99}
\bibinfo{author}{\bibfnamefont{H.-C.} \bibnamefont{Jeong}} \bibnamefont{and}
  \bibinfo{author}{\bibfnamefont{E.~D.} \bibnamefont{Williams}},
  \bibinfo{journal}{Surf. Sci. Rep.} \textbf{\bibinfo{volume}{34}},
  \bibinfo{pages}{171} (\bibinfo{year}{1999}).

\bibitem[{\citenamefont{Politi et~al.}(2000)\citenamefont{Politi, Grenet,
  Marty, Ponchet, and Villain}}]{politi00}
\bibinfo{author}{\bibfnamefont{P.}~\bibnamefont{Politi}},
  \bibinfo{author}{\bibfnamefont{G.}~\bibnamefont{Grenet}},
  \bibinfo{author}{\bibfnamefont{A.}~\bibnamefont{Marty}},
  \bibinfo{author}{\bibfnamefont{A.}~\bibnamefont{Ponchet}}, \bibnamefont{and}
  \bibinfo{author}{\bibfnamefont{J.}~\bibnamefont{Villain}},
  \bibinfo{journal}{Phys. Rep.} \textbf{\bibinfo{volume}{324}},
  \bibinfo{pages}{271} (\bibinfo{year}{2000}).

\bibitem[{\citenamefont{Yagi et~al.}(2001)\citenamefont{Yagi, Minoda, and
  Degawa}}]{yagi01}
\bibinfo{author}{\bibfnamefont{K.}~\bibnamefont{Yagi}},
  \bibinfo{author}{\bibfnamefont{H.}~\bibnamefont{Minoda}}, \bibnamefont{and}
  \bibinfo{author}{\bibfnamefont{M.}~\bibnamefont{Degawa}},
  \bibinfo{journal}{Surf. Sci. Rep.} \textbf{\bibinfo{volume}{43}},
  \bibinfo{pages}{45} (\bibinfo{year}{2001}).

\bibitem[{\citenamefont{Stangl et~al.}(2004)\citenamefont{Stangl, Holy, and
  Bauer}}]{stangl04}
\bibinfo{author}{\bibfnamefont{J.}~\bibnamefont{Stangl}},
  \bibinfo{author}{\bibfnamefont{V.}~\bibnamefont{Holy}}, \bibnamefont{and}
  \bibinfo{author}{\bibfnamefont{G.}~\bibnamefont{Bauer}},
  \bibinfo{journal}{Rev. Mod. Phys.} \textbf{\bibinfo{volume}{76}},
  \bibinfo{pages}{725} (\bibinfo{year}{2004}).

\bibitem[{\citenamefont{Liu. and Weeks}(1998)}]{liu98bis}
\bibinfo{author}{\bibfnamefont{D.-J.} \bibnamefont{Liu.}} \bibnamefont{and}
  \bibinfo{author}{\bibfnamefont{J.~D.} \bibnamefont{Weeks}},
  \bibinfo{journal}{Phys. Rev. B} \textbf{\bibinfo{volume}{57}},
  \bibinfo{pages}{14891} (\bibinfo{year}{1998}).

\bibitem[{\citenamefont{Frisch and Verga}(2005)}]{frisch05}
\bibinfo{author}{\bibfnamefont{T.}~\bibnamefont{Frisch}} \bibnamefont{and}
  \bibinfo{author}{\bibfnamefont{A.}~\bibnamefont{Verga}},
  \bibinfo{journal}{Phys. Rev. Lett.} \textbf{\bibinfo{volume}{94}},
  \bibinfo{pages}{226102} (\bibinfo{year}{2005}).

\bibitem[{\citenamefont{Sato et~al.}(2005)\citenamefont{Sato, Uwaha, and
  Saito}}]{sato05}
\bibinfo{author}{\bibfnamefont{M.}~\bibnamefont{Sato}},
  \bibinfo{author}{\bibfnamefont{M.}~\bibnamefont{Uwaha}}, \bibnamefont{and}
  \bibinfo{author}{\bibfnamefont{Y.}~\bibnamefont{Saito}},
  \bibinfo{journal}{Phy. Rev. B} \textbf{\bibinfo{volume}{72}},
  \bibinfo{pages}{045401} (\bibinfo{year}{2005}).

\bibitem[{\citenamefont{Pierre-Louis}(2006)}]{pierre-louis06}
\bibinfo{author}{\bibfnamefont{O.}~\bibnamefont{Pierre-Louis}},
  \bibinfo{journal}{Phys. Rev. Lett.} \textbf{\bibinfo{volume}{96}},
  \bibinfo{pages}{135901} (\bibinfo{year}{2006}).

\bibitem[{\citenamefont{Dufay et~al.}(2007)\citenamefont{Dufay, Debierre, and
  Frisch}}]{dufay07}
\bibinfo{author}{\bibfnamefont{M.}~\bibnamefont{Dufay}},
  \bibinfo{author}{\bibfnamefont{J.~M.} \bibnamefont{Debierre}},
  \bibnamefont{and} \bibinfo{author}{\bibfnamefont{T.}~\bibnamefont{Frisch}},
  \bibinfo{journal}{Phys. Rev. B} \textbf{\bibinfo{volume}{75}},
  \bibinfo{pages}{045413} (\bibinfo{year}{2007}).

\bibitem[{\citenamefont{Fruchart}(2005)}]{fruchart05}
\bibinfo{author}{\bibfnamefont{O.}~\bibnamefont{Fruchart}},
  \bibinfo{journal}{Comptes Rendus Physique} \textbf{\bibinfo{volume}{6}},
  \bibinfo{pages}{1} (\bibinfo{year}{2005}).

\bibitem[{\citenamefont{Latyshev et~al.}(1989)\citenamefont{Latyshev, Aseev,
  Krasilnikov, and Stenin}}]{latyshev89}
\bibinfo{author}{\bibfnamefont{A.~V.} \bibnamefont{Latyshev}},
  \bibinfo{author}{\bibfnamefont{A.~L.} \bibnamefont{Aseev}},
  \bibinfo{author}{\bibfnamefont{A.~B.} \bibnamefont{Krasilnikov}},
  \bibnamefont{and} \bibinfo{author}{\bibfnamefont{S.~I.}
  \bibnamefont{Stenin}}, \bibinfo{journal}{Surf. Sci.}
  \textbf{\bibinfo{volume}{213}}, \bibinfo{pages}{157} (\bibinfo{year}{1989}).

\bibitem[{\citenamefont{Stoyanov}(1991)}]{stoyanov91}
\bibinfo{author}{\bibfnamefont{S.}~\bibnamefont{Stoyanov}},
  \bibinfo{journal}{Jpn. J. Appl. Phys.} \textbf{\bibinfo{volume}{30}},
  \bibinfo{pages}{1} (\bibinfo{year}{1991}).

\bibitem[{\citenamefont{M\'etois and Stoyanov}(1999)}]{metois99}
\bibinfo{author}{\bibfnamefont{J.~J.} \bibnamefont{M\'etois}} \bibnamefont{and}
  \bibinfo{author}{\bibfnamefont{S.}~\bibnamefont{Stoyanov}},
  \bibinfo{journal}{Surf. Sci.} \textbf{\bibinfo{volume}{440}},
  \bibinfo{pages}{407} (\bibinfo{year}{1999}).

\bibitem[{\citenamefont{Fujita et~al.}(1999)\citenamefont{Fujita, Ichikawa, and
  Stoyanov}}]{fujita99}
\bibinfo{author}{\bibfnamefont{K.}~\bibnamefont{Fujita}},
  \bibinfo{author}{\bibfnamefont{M.}~\bibnamefont{Ichikawa}}, \bibnamefont{and}
  \bibinfo{author}{\bibfnamefont{S.~S.} \bibnamefont{Stoyanov}},
  \bibinfo{journal}{Phys. Rev. B} \textbf{\bibinfo{volume}{60}},
  \bibinfo{pages}{16006} (\bibinfo{year}{1999}).

\bibitem[{\citenamefont{Gibbons et~al.}(2005)\citenamefont{Gibbons, Noffsinger,
  and Pelz}}]{gibbons05}
\bibinfo{author}{\bibfnamefont{B.~J.} \bibnamefont{Gibbons}},
  \bibinfo{author}{\bibfnamefont{J.}~\bibnamefont{Noffsinger}},
  \bibnamefont{and} \bibinfo{author}{\bibfnamefont{J.~P.} \bibnamefont{Pelz}},
  \bibinfo{journal}{Surf. Sci.} \textbf{\bibinfo{volume}{575}},
  \bibinfo{pages}{L51} (\bibinfo{year}{2005}).

\bibitem[{\citenamefont{Gibbons et~al.}(2006)\citenamefont{Gibbons, Schaepe,
  and Pelz}}]{gibbons06}
\bibinfo{author}{\bibfnamefont{B.~J.} \bibnamefont{Gibbons}},
  \bibinfo{author}{\bibfnamefont{S.}~\bibnamefont{Schaepe}}, \bibnamefont{and}
  \bibinfo{author}{\bibfnamefont{J.~P.} \bibnamefont{Pelz}},
  \bibinfo{journal}{Surf. Sci.} \textbf{\bibinfo{volume}{600}},
  \bibinfo{pages}{2417} (\bibinfo{year}{2006}).

\bibitem[{\citenamefont{Pierre-Louis}(2003)}]{pierre-louis03}
\bibinfo{author}{\bibfnamefont{O.}~\bibnamefont{Pierre-Louis}},
  \bibinfo{journal}{Surf. Sci.} \textbf{\bibinfo{volume}{529}},
  \bibinfo{pages}{114} (\bibinfo{year}{2003}).

\bibitem[{\citenamefont{Pierre-Louis and M\'etois}(2004)}]{pierre-louis04}
\bibinfo{author}{\bibfnamefont{O.}~\bibnamefont{Pierre-Louis}}
  \bibnamefont{and} \bibinfo{author}{\bibfnamefont{J.-J.}
  \bibnamefont{M\'etois}}, \bibinfo{journal}{Phys. Rev. Lett.}
  \textbf{\bibinfo{volume}{93}}, \bibinfo{pages}{165901}
  (\bibinfo{year}{2004}).

\bibitem[{\citenamefont{{Zhao} et~al.}(2004)\citenamefont{{Zhao}, {Weeks}, and
  {Kandel }}}]{zhao04}
\bibinfo{author}{\bibfnamefont{T.}~\bibnamefont{{Zhao}}},
  \bibinfo{author}{\bibfnamefont{J.~D.} \bibnamefont{{Weeks}}},
  \bibnamefont{and} \bibinfo{author}{\bibfnamefont{D.}~\bibnamefont{{Kandel
  }}}, \bibinfo{journal}{Phys. Rev. B} \textbf{\bibinfo{volume}{70}},
  \bibinfo{pages}{161303} (\bibinfo{year}{2004}).

\bibitem[{\citenamefont{Degawa et~al.}(2000)\citenamefont{Degawa, Minoda,
  Tanishiro, and Yagi}}]{degawa00}
\bibinfo{author}{\bibfnamefont{M.}~\bibnamefont{Degawa}},
  \bibinfo{author}{\bibfnamefont{H.}~\bibnamefont{Minoda}},
  \bibinfo{author}{\bibfnamefont{Y.}~\bibnamefont{Tanishiro}},
  \bibnamefont{and} \bibinfo{author}{\bibfnamefont{K.}~\bibnamefont{Yagi}},
  \bibinfo{journal}{Surf. Sci. Lett.} \textbf{\bibinfo{volume}{461}},
  \bibinfo{pages}{L528} (\bibinfo{year}{2000}).

\bibitem[{\citenamefont{Burton et~al.}(1951)\citenamefont{Burton, Cabrera, and
  Frank}}]{Burton51}
\bibinfo{author}{\bibfnamefont{W.~K.} \bibnamefont{Burton}},
  \bibinfo{author}{\bibfnamefont{N.}~\bibnamefont{Cabrera}}, \bibnamefont{and}
  \bibinfo{author}{\bibfnamefont{F.~C.} \bibnamefont{Frank}},
  \bibinfo{journal}{Philos. Trans. R. Soc. London, Ser. A}
  \textbf{\bibinfo{volume}{243}}, \bibinfo{pages}{299} (\bibinfo{year}{1951}).

\bibitem[{\citenamefont{M\H{u}ller and Sa\'ul}(2004)}]{muller04}
\bibinfo{author}{\bibfnamefont{P.}~\bibnamefont{M\H{u}ller}} \bibnamefont{and}
  \bibinfo{author}{\bibfnamefont{A.}~\bibnamefont{Sa\'ul}},
  \bibinfo{journal}{Surf. Sci. Rep.} \textbf{\bibinfo{volume}{54}},
  \bibinfo{pages}{593} (\bibinfo{year}{2004}).

\bibitem[{\citenamefont{Krug et~al.}(2005)\citenamefont{Krug, Tonchev,
  Stoyanov, and Pimpinelli}}]{krug05}
\bibinfo{author}{\bibfnamefont{J.}~\bibnamefont{Krug}},
  \bibinfo{author}{\bibfnamefont{V.}~\bibnamefont{Tonchev}},
  \bibinfo{author}{\bibfnamefont{S.}~\bibnamefont{Stoyanov}}, \bibnamefont{and}
  \bibinfo{author}{\bibfnamefont{A.}~\bibnamefont{Pimpinelli}},
  \bibinfo{journal}{Phys. Rev. B} \textbf{\bibinfo{volume}{71}},
  \bibinfo{pages}{045412} (\bibinfo{year}{2005}).

\bibitem[{\citenamefont{{Popkov} and {Krug}}(2006)}]{popkov06}
\bibinfo{author}{\bibfnamefont{V.}~\bibnamefont{{Popkov}}} \bibnamefont{and}
  \bibinfo{author}{\bibfnamefont{J.}~\bibnamefont{{Krug}}},
  \bibinfo{journal}{Phys. Rev. B} \textbf{\bibinfo{volume}{73}},
  \bibinfo{pages}{235430} (\bibinfo{year}{2006}).

\bibitem[{\citenamefont{Chang et~al.}(2006)\citenamefont{Chang, Pierre-Louis,
  and Misbah}}]{chang06}
\bibinfo{author}{\bibfnamefont{J.}~\bibnamefont{Chang}},
  \bibinfo{author}{\bibfnamefont{O.}~\bibnamefont{Pierre-Louis}},
  \bibnamefont{and} \bibinfo{author}{\bibfnamefont{C.}~\bibnamefont{Misbah}},
  \bibinfo{journal}{Phys. Rev. Lett.} \textbf{\bibinfo{volume}{96}},
  \bibinfo{pages}{195901} (\bibinfo{year}{2006}).

\end{thebibliography}


\end{document}